# Reconstructing the Universe in a computer: physical understanding in the digital age

Simon D.M. White
Max Planck Institute for Astrophysics

The influence of Aristotle on the development of academic and intellectual life was immense. He thought and wrote about an extraordinarily broad range of topics, and his ideas on logic, physics, biology, philosophy, ethics, politics, poetics and rhetoric continued to play a major role in European intellectual discourse for nearly two thousand years after his death. Even by the beginning of the seventeenth century, the idea that experimental evidence should underlie physical reasoning was still not fully established, and both Stevinus and Galileo felt it necessary to discuss dropping unequal weights from a tower in order to dispute Aristotle's authority, specifically his statement that objects of differing weight fall at differing speeds. Before the century was over, however, Newton had found a single law explaining how gravity regulates not only the falling of terrestrial bodies, but also the motion of the Moon around the Earth and that of the planets around the Sun. His analytic approach to optics and mechanics laid the basis for the scientific method for the next 400 years – careful empirical characterisation of the properties and behaviour of some aspect of "reality" is followed by the framing of an often but not always mathematical description to be tested against further experiment or observation.

In the last 50 years the unprecedented growth of computing power has brought a new ingredient to the scientific method. Numerical simulation has become the tool of choice for studying complex systems such as forming galaxies or planets, the changing climate, evolving ecosystems, or the global economy. The equations and algorithms underlying such simulations are mathematical idealisations which encapsulate hypotheses (or "models") for various aspects of the behaviour of the real systems. They interact in such a way that a simulation's outcome cannot be predicted in advance, turning it into a numerical "experiment" that explores the behaviour of the set of models rather than that of any "real" system. The scientific challenge lies in assessing whether regularities seen in results of such simulations give insight into the behaviour of the corresponding real system, or merely reflect simplifications inherent in the models. My career in astrophysics began as this revolution was getting underway, and I have been lucky to work in a field where simulations have produced fundamental new insights. However, as I hope to make clear in this personal and historical essay, the role of computation in the scientific process is a complex one which is still developing rapidly.

When I was a student in Cambridge in the early 1970's, the largest available computer was a brand-new IBM 370 which we programmed by feeding punched cards into remote card-readers. Output came back a day later in the form of large stacks of paper from line-printers. These were the only available means to produce an "image" of a simulation. Within 3 or 4 years, the IBM 370 had been junked. Today's largest computers are about a billion times faster and have a billion times larger memory capacity. In contrast, the world's largest optical telescope in the 1970's was the 20 year-old Hale 200 inch on Mount Palomar, This is still a functioning research facility, and the world's largest optical telescopes currently have about ten times the collecting area and use instruments that are roughly 100 times more efficient than the photographic plates of the 1970's. The explosive growth in computing relative to other necessary technical capabilities is responsible for its greatly increased role in today's astrophysics.

As a student, I was interested in whether the "missing mass" in galaxy clusters is attached to the individual galaxies in proportion to their light or is distributed smoothly between them. It seemed to me that a simulation of cluster formation might point to an answer, so I borrowed a star cluster simulation code from Sverre Aarseth and set up initial conditions corresponding to Gunn and Gott's (1972) model for galaxy cluster formation. I placed 700 "galaxies" at random within a uniformly expanding but gravitationally bound sphere, assigned them masses paralleling the luminosity distribution of galaxies in the Coma cluster, and used the IBM370 to follow the system's evolution. In the G&G model, the galaxy distribution stays uniform and spherical, expanding to a maximum size before collapsing, bouncing and relaxing to dynamical equilibrium. My line-printer pictures told a very different and (to me) unexpected story. Already during the expansion, galaxies clumped into ever large groups and at maximum size the system was in two almost equal pieces, These then fell together, merging to make a realistic looking cluster, a decent facsimile of the observed Coma system.

This was my first experience of how numerical simulations can lead to qualitatively new insights. Today, the hierarchical growth of clusters is taken for granted, but at the time the images were both surprising and compelling. The "experiment" did indeed answer my original question – the concentration of luminous galaxies to the centre of the simulated cluster was much stronger than observed, indicating that the dark matter in real clusters is not attached to galaxies in proportion to their light. Nevertheless, the unexpected result was the most exciting one. It was also a lucky one, because a couple of years later the Einstein satellite made the first X-ray images of galaxy clusters and found systems which seemed to correspond to each of the evolutionary stages of my simulation.

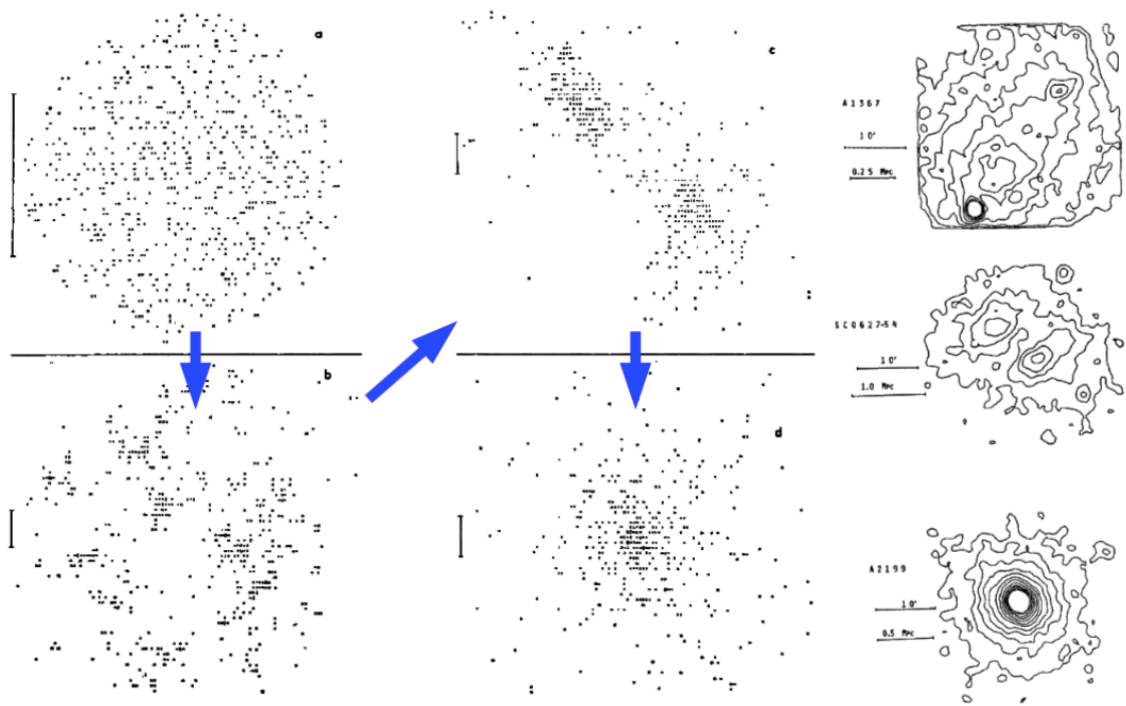

Fig.1 Line-printer images of the evolution of a 700-particle cluster formation simulation from White (1976) are compared with X-ray images of galaxy clusters from Forman & Jones (1982). The (expanding) initial condition for the simulation is at top left and the vertical bar in each simulated image is the same physical length, so that the actual scale of the initial condition is considerably smaller than that of the other images.

Cosmic structure formation is uniquely suited for study through simulations. The initial conditions are simple and can be observed directly in the Cosmic Microwave Background. In addition, they grow into today's large-scale structure primarily through the action of gravity; non-gravitational processes are critical for determining the properties of individual galaxies but have relatively weak effects on larger scales. By the end of 1970's the upper limits on fluctuations in the CMB were already stringent enough to make it doubtful that gravity could produce present-day structure in the time available since last scattering of the CMB photons. A possible solution, that dark matter might be made of neutrinos, was greatly encouraged by a 1980 tritium decay experiment which claimed an electron neutrino mass of 30 eV. A critical question was then whether the growth of structure in a neutrino-dominated universe could be consistent with the large-scale structure seen in the present-day galaxy distribution.

Testing this hypothesis required calculation of the expected (linear) initial conditions at high redshift, an algorithm to impose these IC's on an N-body system, a code capable of evolving forwards to the present day, and a scheme to put "galaxies" into the result in order to compare it with observation. All these requirements were in place by the early 1980's; the last, potentially most difficult, step could be finessed, because the coherence length in neutrino-dominated universes is so large that although it is very uncertain how galaxies should populate the nonlinear filaments and sheets in which they *could* form, it is clear that there should be *no* galaxies in the large low-density regions between them where no nonlinear structure is expected. As Fig.2 shows, the large voids that this produces in the galaxy distribution were incompatible even with the relatively meagre observational data available in 1983. This discrepancy led to the abandoning of the known neutrinos as potential dark matter candidates, even though it would be another two decades before they were finally excluded by experimental upper limits on their masses. The demonstration that no known particle can account for the dark matter remains one of the most significant contributions of computer simulations to astrophysics and cosmology.

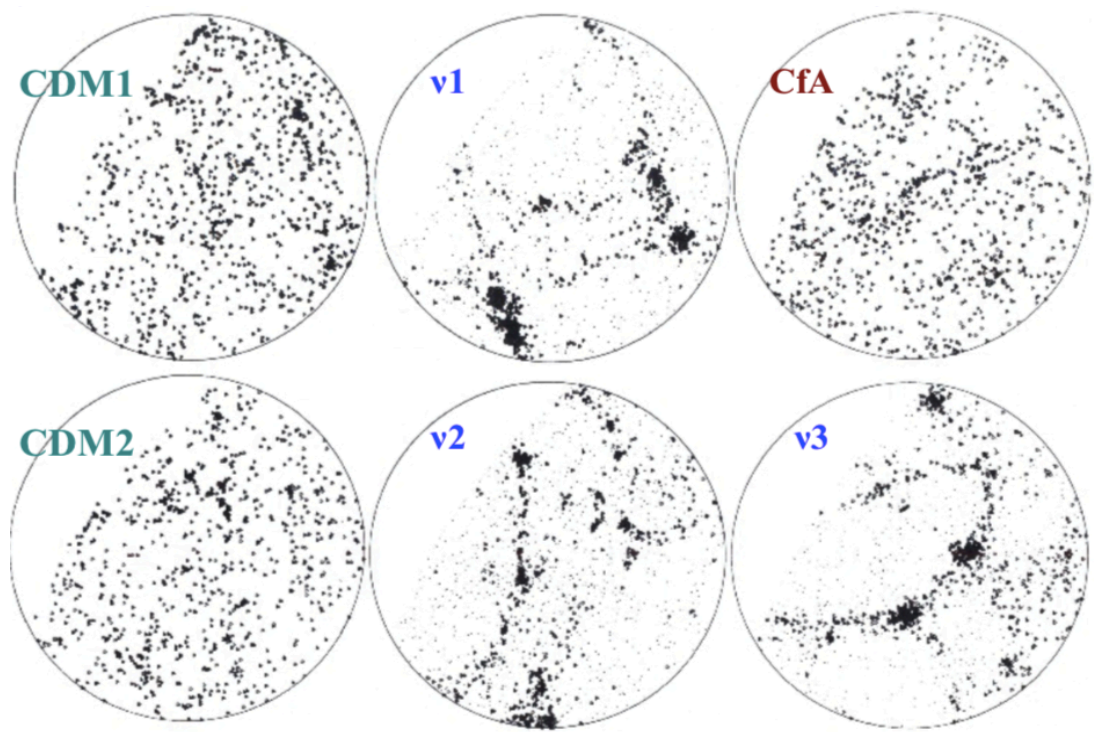

Fig.2 The distribution of intrinsically luminous galaxies brighter than apparent magnitude 14.5 on the northern sky, taken from the CfA survey of Davis et al (1982) is compared with the distribution predicted in neutrino-dominated and cold dark matter dominated universes. Only the large symbols could be visible galaxies in the neutrino case (image taken from White 1986).

The images of Fig.2 also provide an early example of another important use of computer simulations: they make it possible to present theoretical results in a form which allows direct comparison with observation. In particular, they allow relatively direct accounting for the effects produced by observational selection and model simplification, here the geometry and the magnitude limits of the CfA survey and the specific assumptions about how the distribution of galaxies relates to that of dark matter. While the assumption that local nonlinear collapse is required for galaxy formation was sufficient to exclude neutrino-dominated universes, the apparently better agreement of Cold Dark Matter (CDM) universes with observation depends on the assumption underlying the CDM images of Fig.2, that galaxies are distributed like the dark matter, independent of their luminosity. Even in the early 1980's this seemed oversimplified; galaxies were thought to form by the cooling and condensation of gas at the centres of dark matter halos, with more luminous galaxies forming in more massive halos.

A more realistic treatment of this issue requires substantially more powerful simulations so that the formation and structure of dark matter halos can be followed over a wide range of masses throughout a representative region of space. The CDM simulations of Fig.2 used 30,000 particles to track the mass distribution in a cubic volume 14,000 km/s on a side. Twenty years later, improvements in computer power and in integration algorithms made possible the Millennium Simulation (MS) which followed 10,000,000,000 particles in a cube 50,000 km/s on a side. While the increase in volume by a factor of 45 resulted in a substantial improvement in the accuracy with which large-scale structure statistics could be calculated, it was the improvement in mass resolution by a factor of 7,000 which allowed the growth of dark matter halos to be followed in detail, and hence the formation and evolution of the galaxies within them to be tracked by simple, physically based models. By 2005 observational surveys of the galaxy distribution had also improved dramatically, and Fig.3 shows how well galaxy formation simulations based on the MS were able to represent contemporary observations.

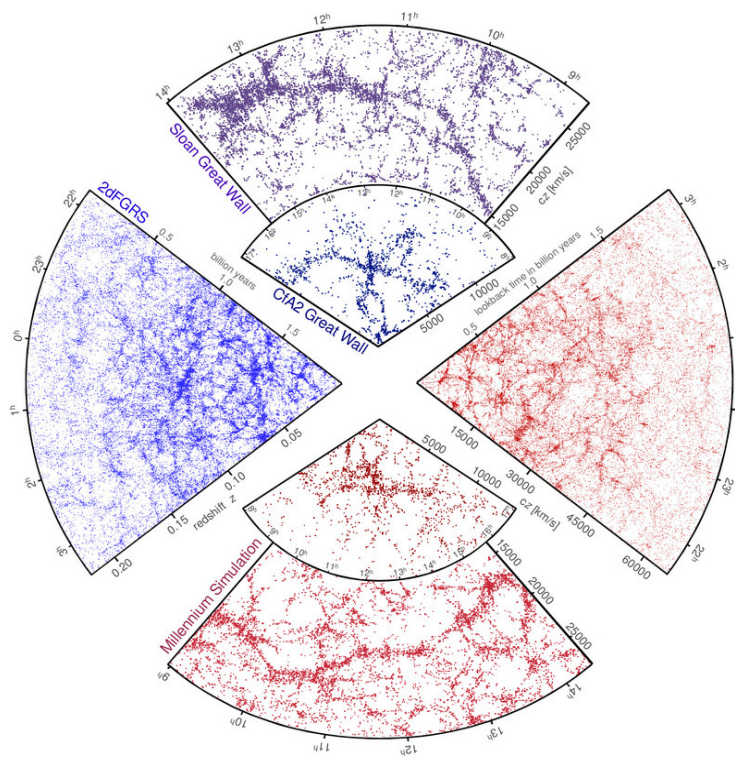

Fig.3 The observed distribution of galaxies in slices through the 2dF (left), SDSS (upper top) and CfA (lower top) surveys is compared with the distribution of galaxies selected using similar geometric, magnitude, colour and redshift criteria applied to a mock catalogue constructed from the Millennium Simulation. When simulating the formation of the galaxies, a number of control parameters associated with star formation and feedback processes were adjusted in order to reproduce the observed abundance of galaxies as a function of stellar mass. Galaxy clustering was not used as a constraint, however (from Springel, Frenk & White 2006).

In addition to the growth and merging of dark matter halos, galaxy formation involves many processes associated with the cooling and condensation of gas, with star formation and evolution, with galaxy merging, with the formation of central supermassive black holes, and with feedback on surrounding material induced by all these processes. Simulations like that of Fig.3 show that the observed abundance of galaxies as a function of luminosity cannot be reproduced within the dark halo population which forms in a CDM universe unless feedback strongly supresses star formation both at low mass, where it can be ascribed to energy input from massive stars and supernovae, and at high mass, where it is plausibly but less certainly associated with Active Galactic Nuclei (AGN). Demonstrating that feedback is required if the observed galaxy population is to be consistent with the initial fluctuations and the proportions of dark matter, baryons and dark energy inferred from the CMB is one of the major contributions of simulations to physical cosmology. However, the schematic and parametrised way that feedback is treated in the model of Fig.3 means that the details of the mechanisms, and in some cases even their true nature, remain obscure; the fit to observations determines at best a few overall efficiencies and scaling exponents. Nevertheless, it seems significant that a model with parameters chosen to fit present-day galaxy abundances can also fit present-day clustering, as well as the evolution of both abundances and clustering out to high redshift.

Fig.4 shows an example of how such simulations can be used to test aspects of the cosmological model which are very different from those used to set their parameters. Gravitational lensing data from SDSS are here compared to an MS-based simulation similar to that of Fig.3 but using an updated cosmology together with galaxy formation parameters that exactly reproduce observed galaxy abundances as a function of stellar mass. Galaxies at the centres of their dark matter halos were identified in a similar way in the observed and simulated catalogues, and mean mass density profiles around them were calculated as a function of the galaxy's stellar mass. Although neither gravitational lensing nor galaxy clustering observations were used to constrain the galaxy formation model, and no additional free parameters were adjusted for the purposes of Fig.4, agreement is good over a factor of 1,000 in projected radius and for galaxies with stellar masses ranging from a third that of the Milky Way to those of the central galaxies of rich clusters. This suggests that the dark matter surrounding galaxies must be the same as that present at last scattering, and that its structure has grown from that observed in the CMB by dissipationless gravitational clustering as in the standard LCDM cosmology.

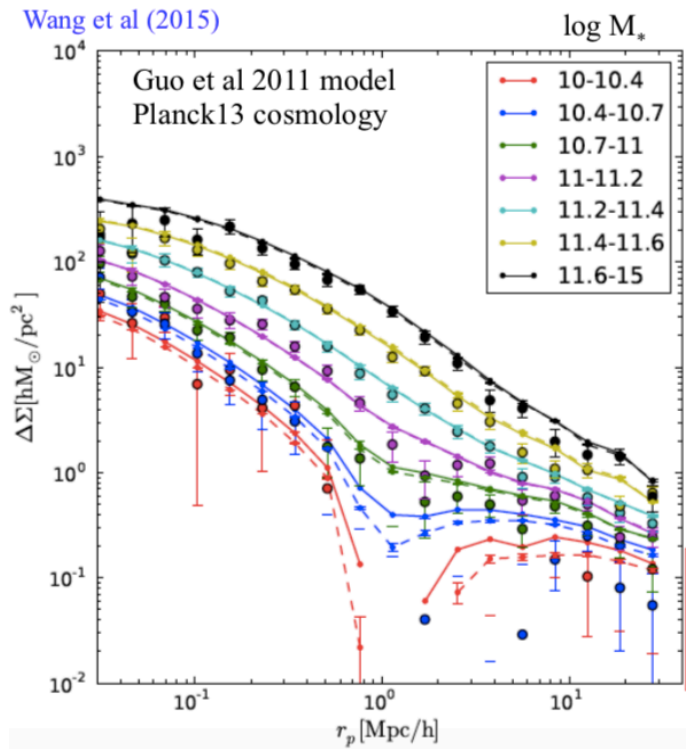

Fig.4 Mean mass surface density profiles around observed galaxies are compared to those in a simulation with galaxy formation parameters tuned to reproduce the observed galaxy abundance in a *Planck* cosmology. The observational results are shown as symbols with error bars and are based on SDSS data for gravitationally lensed distant galaxies behind a large sample of nearer systems. Results are plotted as a function of projected radius for different values of the stellar mass of the central galaxy. Solid lines show simulation results for the same bins in central galaxy stellar mass. On small scales the signal is due to the halo of the central galaxy itself, but on large scales it is due primarily to other halos clustered around the central one.

Another important application of simulations is in uncovering regularities in the behaviour of nonlinear systems which can be accurately reproduced by tuning a suitable phenomenological model. Such a model can provide intuitive understanding of the systems, allowing them to be represented in contexts where direct simulation would be prohibitively expensive. The best example is undoubtedly the Navarro-Frenk-White formula for the spherically averaged density profiles of dark halos. Although the NFW profile has only two parameters, a characteristic radius and a characteristic density, it provides a good fit to the mean profile of simulated halos of given mass over a very wide range in radius and halo mass. Furthermore, the fit is good for cosmologies with a much broader range of parameters than are consistent with current observations, and the relation between the two NFW parameters (which can be recast as halo mass and central concentration) depends on cosmology and the nature of the dark matter in a way which can be understood intuitively as reflecting halo growth histories. This offers a way to constrain both cosmology and the nature of dark matter observationally, for example, by using gravitational lensing as in Fig.4. The accuracy with which this can be done is illustrated in Fig.5

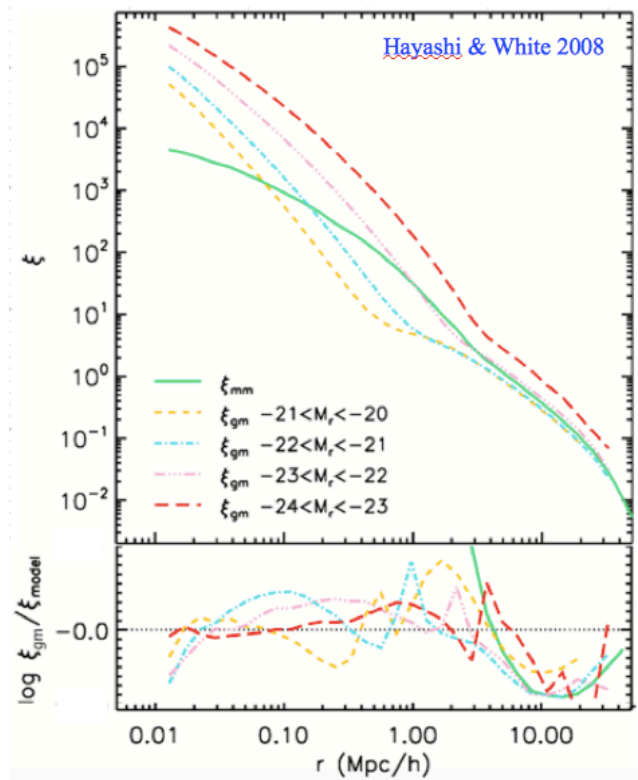

Fig.5 Mean spherically averaged mass density profiles around the central galaxies of dark matter halos as a function of the absolute magnitude of the central galaxy. The top panel shows results from a simulation similar to those used to make Figures 3 and 4. The vertical axis is given in units of the mean density of the universe. The kinks in these profiles reflect the transition between domination by the galaxy's own halo and by the halos of its neighbours. The lower panel shows the density ratio of these directly measured profiles to an analytic model where halo abundances, mean density profiles and clustering bias are all given by simple cosmology-dependent phenomenological models. One tick on the vertical axis here marks 0.01 dex.

This test compares the mass profiles measured around galaxies in a very large cosmological simulation to predictions made by combining three different analytic models. These specify the abundance of halos, their individual density profiles and their spatial distribution, all as a function of halo mass and cosmology. For each model, the adjustable parameters of an empirically or theoretically derived analytic form were tuned to fit the results of earlier simulations. The model combination fits the measured mass profiles to within about 10% for central galaxies of all luminosities and over more than 3 orders of magnitude in radius, corresponding to about 7 orders of magnitude in density. This is good enough to predict lensing signals of the kind shown in Fig.4 quite accurately, and so to explore their sensitivity to cosmological parameters. The utility to the community of such simulation-calibrated analytic models is evident in the fact that our article setting out the NFW profile and exploring the cosmology and halo mass dependence of its parameters is currently the most highly cited paper in theoretical astrophysics of the last thirty years, and the most highly cited paper ever on astrophysical simulations.

In recent years the focus of activity among researchers interested in simulating cosmological structure formation has shifted away from the N-body methods which underlie the simulations I have discussed so far towards methods which explicitly follow the evolution of the baryonic components. Diffuse gas is treated with hydrodynamics or MHD codes and this gas is turned into stars or supermassive black holes (SMBHs) in regions where conditions are appropriate. Evolutionary models for these stars and SMBHs then provide recipes for the radiative, hydro-dynamical and material feedback which they inject into their surroundings. Such schemes provide a detailed representation of galaxy formation, and so permit a much closer comparison with observation. During the first fifteen years of such work, it proved difficult to create objects resembling real galaxies like our own, but in the last decade improved numerical techniques, more powerful computers and accumulated experience in representing feedback have enabled various groups to carry out simulations of galaxy formation in a LCDM universe which produce quite convincing look-alikes for the Milky Way. An example is shown in Fig.6.

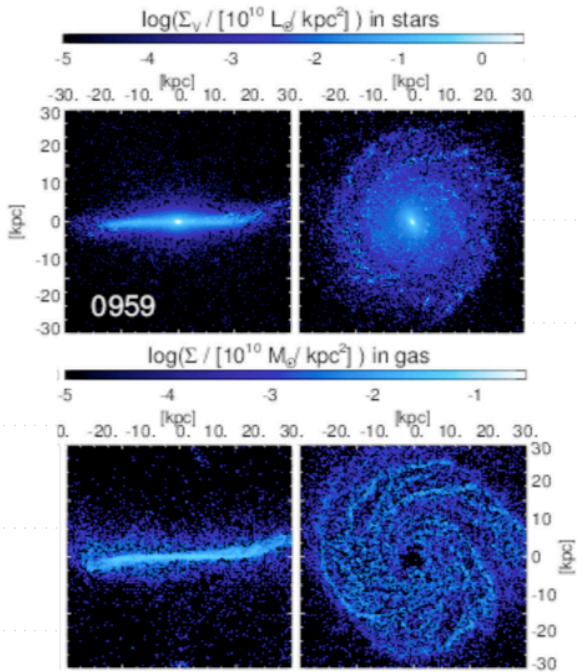

Fig.6  The present-day object which forms at the centre of a halo with mass similar to that of the Milky Way in a simulation from LCDM initial conditions. The upper panels are edge-on and face-on visual band images of the stars, while the lower panels show images of the gas surface density. The simulation used a multi-phase hydrodynamics scheme to follow the gas. It assumed that stars form when the gas reaches high density, and that they drive feedback into surrounding hot and cold gas as they age. As in the Milky Way, most of the stars are in a thin disk that has spiral arms a and clear warp, while there is a small bar and a high surface density bulge at the centre, together with a gas density minimum. The elemental abundances in the stars and the gas also resemble those observed in the Milky Way (taken from Aumer et al 2013).

Such simulations show that it is plausible for galaxies like our own to form in a universe with the properties inferred from the CMB and from low-redshift large-scale structure. This is reinforced by even more recent hydrodynamic simulations which follow galaxy formation throughout "representative" regions. Although much smaller than the region covered by the 10 years older but more schematic simulation of Fig.3, these are still large enough to show that the abundances, structural properties and clustering of galaxies can all be reproduced once suitable recipes are chosen for unresolved small-scale processes such as star formation and feedback. However, the results are very sensitive to the details of how these processes are represented, and agreement with observation requires careful tuning of the particular phenomenological models chosen. Thus, while recent simulations eliminate many of the crude models for baryonic processes which underlie older work and enable much more detailed comparison with observed galaxy structure, they still have to be fitted to observation by adjusting the parameters of *ad hoc* models for unresolved processes. This situation seems unlikely to change soon, given that there is ample evidence that the galaxy population is strongly influenced by feedback associated with AGN at the centres of massive galaxies.

This highlights a significant methodological issue. Galaxies are complex and diverse systems with many distinct components and can be characterised observationally using a great variety of techniques. As a population they exhibit regularities in the form of relations between properties such as mass, size, morphology, gas content, chemical composition and environment, but these usually show substantial scatter and there is often an interesting subpopulation of outliers. From a theoretical point of view, it is clear that many different processes play a major role in their formation and evolution, for example, gas cooling and condensation, star formation and stellar evolution, the formation of central SMBHs, radiative, hydrodynamic and material feedback from stars and AGN, chemical enrichment and galaxy mergers. There is no agreement about the completeness of this list – e.g. magnetic fields and the relativistic "cosmic ray" component may play a role in driving galactic winds and are clearly important for feedback at the centre of galaxy clusters. All these processes are strongly coupled, so that studies of their individual effects can tell at most a partial story. A consequence is that simulations must represent all of them if they are to produce a "realistic" galaxy population. Furthermore deficiencies in the phenomenological treatment of unresolved processes can significantly affect larger scales that may appear well resolved. For example. schematic treatments of wind generation by supernovae and AGN do not reliably predict the phase structure of the ejected material and hence the way it mixes with circumgalactic gas. This leads to uncertainties in the rate at which galaxies grow as diffuse material cools and condenses from their surroundings.

The issue, then, is how to gain understanding of complex systems by analysing simulations where unresolved processes that substantially affect the global behaviour are represented by schematic and parametrised "subgrid" models. While it is encouraging that the simulation of Fig.6 matches many properties of the Milky Way quite closely, the structure of the interstellar medium is not properly represented on the scales that control star formation and stellar feedback, and the stellar mass, morphology, size and gas content of the final galaxy all depend sensitively both on the form and on the parameters of the subgrid models adopted for these processes. Thus it is unclear how the apparent success of the simulation should be interpreted. For the more schematic galaxy formation models of Figs 3, MCMC or equivalent methods allow exploration of the full parameter space of the subgrid models, and hence evaluation of the constraints posed by observation, but this is not computationally feasible for hydrodynamic simulations which must be repeated in full for every change in subgrid model or parameters. In my view, it is not enough to show that one's favorite simulation produces a galaxy or a galaxy population in good agreement with (certain aspects of) observed reality; this may simply be a consequence of limitations in the subgrid modelling. Progress requires physical understanding of *why* some simulations fit observation while others with different parameters or different treatments of the astrophysics do not. For a system as complex and multi-facetted as the galaxy population, such understanding can only be gained through analysis of many simulations where subgrid models are varied over the full range allowed by physical consistency.

The difficulty in using complex simulations of limited (and ultimately unknown) fidelity to gain insight into the behaviour of real systems of even greater complexity is not limited to astrophysics, but is present in many other fields, for example, climate change, ecology, brain function and macro-economics. As the power of computers and their use for numerical experimentation continue to explode, new and systematic epistemological strategies will need to be devised in order to distil useful and lasting scientific knowledge from their results.